# Smart city analysis using spatial data and predicting the sustainability

Thomas Joseph[1]

[1]Asst, Professor, Department of Computer Science and Engineering,
St.Joseph's College of Engineering and Technology, Palai, Kottayam, Kerala.

*Abstract*— Smart city [1] planning is crucial as it should balance among resources and the needs of the city .It allows to achieve good eco-friendly industries, there by supporting both the nature and the stake holders. Setting up an industry is a difficult problem, because it should optimize the resources and allocating it in an effective manner. Weighted sum approach [2] uses the spatial data for finding appropriate places to set up the industry based on the weight assigned to each constraint. The user can predict the possible places in the search space, where the industry can be set with low time complexity using spatial data. Diversity being introduced by using multipoint crossover and mutation operations. It will help to bring exploration in the search space, thereby bring the diversity factor into the solution space. The prediction approach will help to avoid the human exploitation on nature for resources. This in turn helps the investors to maximize the Return on Investment (ROI).

## I. INTRODUCTION

**IN** less than 40 years, 70% of the world's population will reside in cities. This rapid migration will push both current and future urban centers to maximize and expand the industrial, residential and infrastructures beyond their breaking points. There should be an integration among wider smart city [1] with operational and informational efficiency. An eco friendly city with the best suited business models helps to make efficient infrastructure for the government, investors and employees. It will help the investors to convert the capital expenditure to operational expenditure.

Traditionally transportation, power, communication facilities, labour, population, emergency response, buildings, hospitals, and public services systems of a city operated separately. A truly efficient city requires not only operating in an integrated manner, but the resources for each system should be optimized .It helps the systems managed in an integrated way for better prioritize investment and to maximize value of return. The effects of the system after get implemented need to be tracked, which cannot be tracked with the existing methodology. An integration among constrains is sufficient for a sustainable smart city [1].

The main objective is to develop, provide suitable analysis and suggestions for smart cities using dimensions such as mobility, energy, communication facilities, labour and population. The input data are extracted from spatial data. Spatial data of vector data format is the dataset used for the application. The vector data set is been chosen because it will have more significant values of that place when compared to the scalar data set format

A smart city [1] prediction is a list of prioritized business drivers and technology trends that we believe will shape the local government industrial landscape in the context of an economically and socially challenging time. The goal is to reshaping the ways cities operate and require the city leaders to make smart and to take very difficult decisions about the dimensions social infrastructure. What varies is the degree of intelligence, depending on the person, the system of cooperation, and digital infrastructure and tools that a community offers its residents.

In this work, we use the spatial data to identify places based on constrains. The approaches used are weighted sum approach [2] and brute force method which been used to identify the places suitable for setting up the industry. The proposed system helps the industrialist to identify the places where the government can allow to set up the industry .It will indirectly helps both the investors and employees. Investors can maximize the returns, the employees can get a good working atmosphere .It also helps the industrialist to set the industry in the places without exploitation of natural resources.

## II. LITERATURE SURVEY

Smart city [1] need to emerge for the growth of the era. It should decouple the political element for the improved city servicing, by using the underlying technologies. The concept of sustainable city for the further era evolved by using the concept of image processing. It has been evaluated that a rapid movement of people to the banks of the river for better life style and urbanization in China. The image on analysis shows that the growth of the city and deformation of big manmade structures that happened over there, which signs the development rate in the banks of rivers was ongoing in a faster rate. The project was termed 'Dragon' which had joined participation of both European Space Agency and National Remote Sensing Center of China [3].

A city to be smart it should be in an integrated platform were all constrains should be controlled in an integrated manner. The integrated approach always outperforms the centralized approach in all aspects. The approach guaranteed the synergy factor which directly helps to provide an icon for a sustainable industry. For setting up an industry both the hard as well as soft constrains need to be considered. In this prediction approaches user will use hard constrains for setting up the industry.

The major dimensions taken in the approach are mobility, power, communication facilities, labour factors and population. Mobility mainly includes roadway, railway and water ways. The factor mobility is taken in account because it's mainly used for transportation of raw materials or products from one place to another place. Mobility also helps the workers to travel from one place to another place. Power is another major driving factor for industries. The main sources for the power supply are hydro electric power projects, thermal plants, nuclear plants, LPG/diesel plants and wind mills. Communication is another major factor for setting a smart industrial area. It includes wired, wireless and communication densities of that area. The other factors that need to be taken in account are the labour facilities and the population density of that area. A sustainable icon for a smart city [1] should consider soft constrains also which include the tax policy of the government, the legal frame work, leisure places. Based on the above factor the Pareto set is been populated. The set consist of the possible places were the industry can be set.

For populating Pareto set there are many methods random sampling, weighted sum method, distance method and constrained trade off method. The user needs to find the dominated solution from the Pareto optimal solutions. The weighted sum approach [2] will outperform all other approaches, because the basic assumption made in this spatial application is that the user will have the pre historic values for each constrains corresponding to the industries need to set.

### III. SYSTEM DESIGN

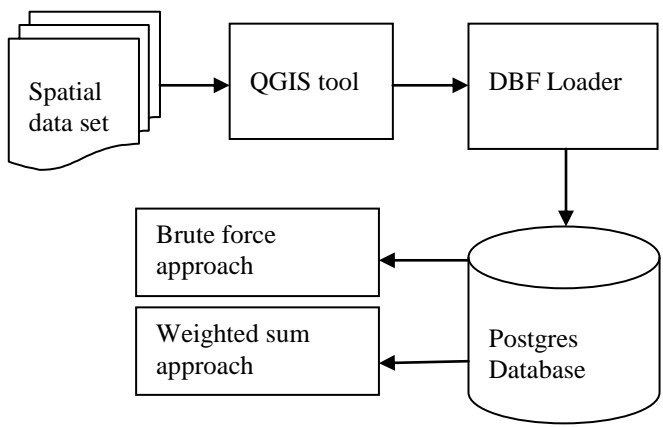

Figure 1: System Architecture

The input to the application is spatial data; it's of the form of a vector data type with the extension .shp files. The sample data set will be loaded in the QGIS. It will be exported to the spatial database by using DBF Loader. The stored data in the data base is then further processed using two approaches (ie weighted method and brute force) for predicting the places, where the industry can be set. The industries thus build based on the analysis of spatial data will be eco-friendly.

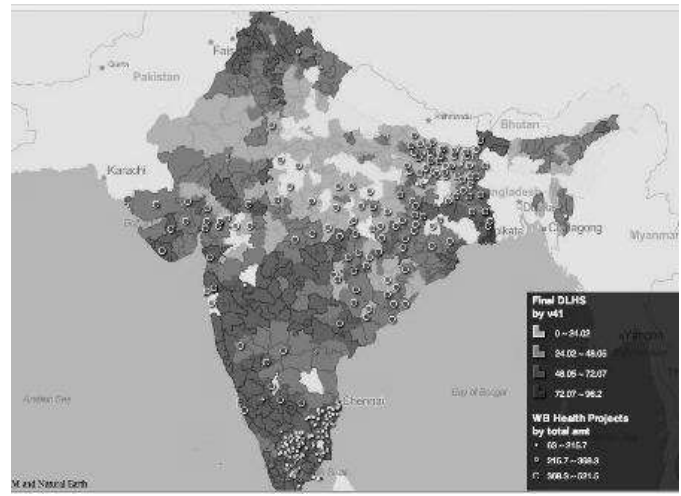

Figure 2: Spatial Data set of India

The above spatial data map of India shows how the health related facilities available in India .The spatial data helps the user to get accurate information, which directly helps to set health centers for people based on the factors of that region. For the experimental work the test data taken is for south India data set.

The use case diagram shows which user will get interact with the modules in the prediction approach. The major modules of the approaches are random selection of point, weight assign to each constrains, converting weight to closed interval, diversity module and fitness evaluation using threshold value.

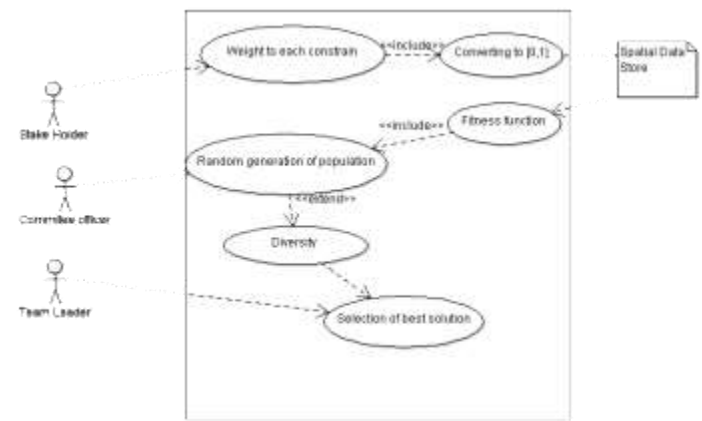

Figure 3: Use case Diagram – Weighted sum approach

The actors in the prediction approaches are stake holders, committee officer and team leader.

### IV. EXPERIMENTAL WORK

The approach mainly uses three methods for extracting and predicting the information from the spatial data .The base model helps to extract the spatial data from the input file. The weighted sum approach works on an integrated approach. Based on the random approach the solution will be processed on the bases of the spatial data set. The brute force approach evaluates all the search space in the domain. Most of the time the brute force approach guarantee the best solution to the problem compared to the weighted sum. Random generation of points is the bottle neck for the weighted sum approach.

The base model works in a scattered manner where the other two approaches works in a integrated manner which helped to determine the places for sustainable smart city [1].

A. BASE MODEL - Geographical Information System

The base model is implemented using the geotool API. The coded application will help the user to retrieve the features of that particular area. It also help to retrieve the X,Y co-ordinates of the location for easy identification to the user. First the developer has to add the API of geo tool to the application library. After adding the library the necessary packages to the application are added. The main packages are the Layer, Mapcontent, Feature Layer and the File data store. First the user needs to upload the vector data format of the spatial data that need to get processed. If the file is null it will return null value with appropriate error message. If the file uploaded is not null it will process the file. The File data store will store the file formats of the file in the application. The feature source will store the feature information of that particular point. For the fast retrieval of the features to the users there is a caching feature source which helps for the fast retrieval of features to the user.

By using the Mapcontent function the user will be able to retrieve the all the feature information of the point and get popup in a JFrame. It will contain the details in that particular point and displayed to the user. The above displayed output can be stored in any format the user wish for further reference. Usually the user prefers to store the above retrieved file in text format.

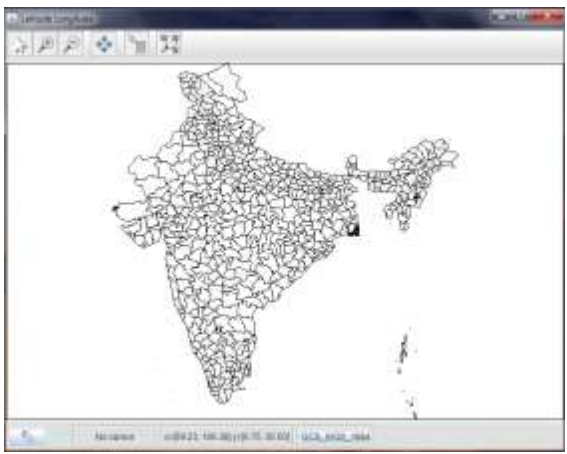

Figure 4: Base model for spatial data analysis

B. WEIGHTED SUM APPROACH

Government and industrialist know the weight that need to allocate to each industry constrain, according to the policy of the government.

The algorithm is a multi-objective evolutionary algorithm which randomly generates an initial population. It then computes the objective constrains for each individual population and store it as separate non-dominated and current population. After apply the fitness function and get the best of the available from the current population. The fitness function is calculated on the basis of the weights assigned to each. The value get computed which get checked to the minimum value that need to get satisfied, if its yes then the iteration get stop else repeats until an optimal solution being got.

ALGORITHM

The weighted sum approach [2] contains many sub modules in it. Initialization of the population is the first sub module in this approach. The population will be generated using the random function available in the java program. If the generated number match with the existing number it will be initialized else again the process of random generation continuous upto the desired population size. The next sub module is the weight assigned module, were the stake holders will assign values to the six constrains that had been taken in account. The weights are denoted by $r_i$ where i= 1,2,3,..m. Specifying weights to each constrains is not easy because the stake holder set a pessimistic target value. The above entered weights is then converted to a [0,1] for better accuracy and accurate result of the threshold value.

$$w_i = r_i \Big/ (r_1 + r_2 + ... + r_m) \quad \text{------} \quad \text{Eq:(1)}$$

Based on the weight assigned to each constrains, the objective function will be get evaluated. The threshold value meet the minimum threshold set by the user those places value will be get store in the remark table in Postgres SQL. The objective function is defined as follows

$$f(x) = w_1 \cdot f_1(x) + w_2 \cdot f_2(x) + ....w_m \cdot f_m(x) \quad \text{------} \quad \text{Eq(2)}$$

If the threshold value of two places does not meet the threshold expected as by the user the diversity module will be get executed. Mutation is done by swapping and crossover uses the multi point crossover technique. The above two approaches in the diversity module will sometime bring diversity factor to the approach. The user who assigns the role as team leader will then evaluate the best places on the basics of the threshold value. The above value can be differ from the minimum threshold that set by the committee team. The value will be based on the pre historic knowledge of the team leader.

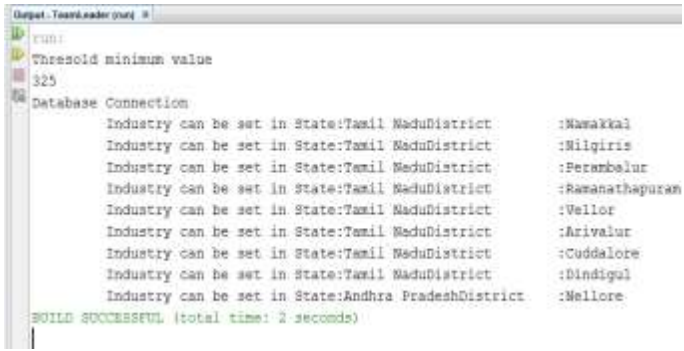

Figure 5: Output Screen of Weighted sum approach

C. BRUTE FORCE

In Weighted sum approach [2] the best solution for the problem may not be guaranteed .Random selection places the bottle neck for the best possible solution in the search space.

In Brute force approach all the places in the search space will be taken in account .The weights that assigned to each constrains will be allocated as above approach .All the places will be get evaluated based on the assigned weights. The best place can be selected based on a threshold value.

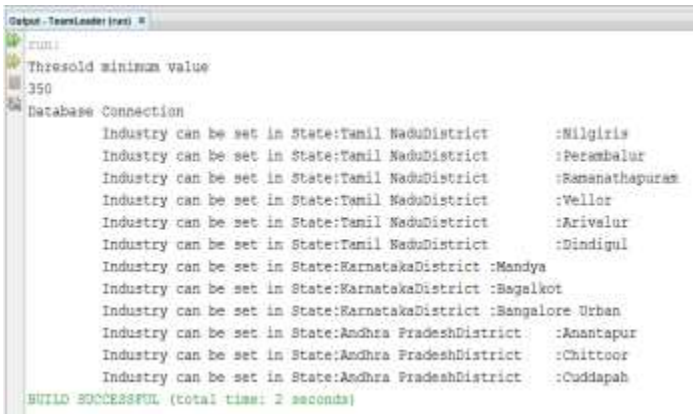

Figure 6: Output Screen of brute force approach

V. RESULTS

Sustainable city with optimal resources allocation was the goal of the spatial application. The base model give solutions to the scattered smart city, while both the other approaches give solutions to the integrated smart city .It combine six major dimensions of the smart city considering the constrains for setting up the industry.

VI. COMPARISON

The brute force and weighted sum approach [2], helps the user to suggest the best possible place to set up the industry.

The weighted sum approach [2] mainly used because simple mathematical model, so the work flow can be easily understand by the users. Its Computational efficiency is suitability to generate a strong solution is another advantage. Time taken for the algorithm will be an upper hand when compared to other multi objective evolutionary algorithm. Approach doesn't generate the best optimal solution in the space because the approach follows the random selection for the solution generation mechanism. The non-convex problem in the approach is another disadvantage regardless of the weight used. The approach does not guarantee the exploration of the search space.

The brute force approach guarantee the best solution in the search space. The approach doesn't use random selection process which indirectly leads to high time complicity for the solution to obtain.

| Factor | Weighted sum | Brute force |
|---|---|---|
| Time complicity | Low | High |
| KLOC | High | Low |
| Diversity factor | Yes | No |
| Solution Search space | Random | All |
| Best feasible solution | Not guaranteed | Yes |

Figure 7: Comparison of weighted sum and brute force

Using the prediction approaches user was able to predict the sustainable places for setting an industry. In weighted sum approach [2] the user uses the random selection process which was the bottle neck for the best optimal result. The weighted sum approach [2] solution may not be the best optimal solution in the search space. To avoid this disadvantage the user use brute force approach, which is time consuming but provides the best optimal solution for the given search space.

VIII. FUTURE ENHANCEMENT

The approaches were developed based on a set of static constrains. So dynamically adding constrains to the previously added constrains. A given scenario were user need to set some specific set of constrains within a group need to be satisfied, cannot be solved by using brute force approach .Solution's for the above are the possible enhancement to the current methodology.

REFERENCES

[1].Ignasi Vilajosana, Jordi Llosa, Borja Martinez, Marc Domingo-Prieto and Albert Angles, "Bootstrapping Smart Cities through a Self-Sustainable Model Based on Big Data Flows", IEEE Communications Magazine June 2013.